\documentclass{emulateapj}
\def\stacksymbols #1#2#3#4{\def\theguybelow{#2}
        \def\verticalposition{\lower#3pt}
        \def\spacingwithinsymbol{\baselineskip0pt\lineskip#4pt}
        \mathrel{\mathpalette\intermediary#1}}
\def\intermediary #1#2{\verticalposition\vbox{\spacingwithinsymbol
        \everycr={}\tabskip0pt
        \halign{$\mathsurround0pt#1\hfil##\hfil$\crcr#2\crcr
                \theguybelow\crcr}}}
\def\lta{\stacksymbols{<}{\sim}{2.5}{.2}}
\def\gta{\stacksymbols{>}{\sim}{3}{.5}}

\shorttitle{RADIATING BONDI FLOWS}
\shortauthors{MATHEWS \& GUO}

\begin{document}

\title{Radiating Bondi and Cooling Site Flows}

\author{William G. Mathews\altaffilmark{1} 
and Fulai Guo\altaffilmark{1}}

\altaffiltext{1}{University of California Observatories/Lick
Observatory,
Department of Astronomy and Astrophysics,
University of California, Santa Cruz, CA 95064
mathews@ucolick.org}

\begin{abstract}
Steady accretion of a radiating gas onto a central 
mass point is described 
and compared to classic Bondi accretion. 
Radiation losses are essential for accretion flows to 
be observed.
Unlike Bondi flows, 
radiating Bondi flows pass through a sonic point at a 
finite radius and become supersonic near the center. 
The morphology of all radiating flows is described 
by a single dimensionless parameter proportional 
to ${\dot M}/MT_s$ where $T_s$ is the gas temperature 
at the sonic point. 
In radiating Bondi flows 
the relationship between the mass accretion rate 
and central mass, ${\dot M} \propto M^p$ 
with $p \sim 1$, 
differs significantly from the quadratic dependence 
in classical Bondi flows, ${\dot M} \propto M^2$. 
Mass accretion rates 
onto galaxy or cluster-centered
black holes estimated from traditional 
and radiating Bondi flows are significantly different.
In radiating Bondi flows the gas temperature 
increases at large radii, as in the cores of many 
galaxy groups and clusters, allowing radiating Bondi 
flows to merge naturally 
with gas arriving from their cluster environments.
Some radiating flows cool completely before reaching 
the center of the flow, 
and this also occurs in cooling site flows, 
in which there is no central gravitating mass. 
\end{abstract}

\vskip.1in
\keywords{hydrodynamics, galaxies: cooling flows, 
galaxies: clusters}

\section{Introduction}
Bondi flow describes the steady spherical 
inflow of an adiabatic gas toward an accreting mass point. 
As gas 
flows radially toward the gravitating mass, 
its density increases due to  
tidal compression by the gravity field and consequently 
its temperature also increases. 
Gas at rest at infinity with uniform temperature and density 
accelerates steadily 
as it flows inward toward the central mass.
In Bondi flow the mass of flowing gas is assumed to be 
small so its self-gravity can be neglected 
and the central mass can be 
regarded as constant during relevant flow times (Bondi 1952).
Conventional non-radiating Bondi flows are often used
in computational studies of massive black hole growth
during galaxy formation or to detect errors in hydrocodes.
Bondi flows are also often used to estimate  
black hole accretion rates from X-ray observations of
gas near central black holes in galaxies and galaxy clusters.
However, Bondi flows do not radiate.

For more realistic astrophysical applications, 
such as the flow of hot intracluster gas 
toward a cluster-centered black hole,
the gas must radiate as it flows inward. 
Our purpose here is to illustrate that steady 
spherical accretion flow solutions  
with radiative losses differ 
fundamentally from the traditional Bondi solution.
A large family of morphologically varied
radiating solutions are possible, all characterized
by a single dimensionless parameter. 
But the most important difference is that the 
often-assumed quadratic Bondi relation between 
the mass accretion rate and the central mass, 
${\dot M} \propto M^2$, no longer holds.
Instead, radiating flows require 
${\dot M} \propto M^p$ with $p \sim 1$. 
Unlike classical Bondi flow, 
the gas temperature in radiating solutions 
always increases at large radii,
allowing more natural fits to the 
hot gas cores in galaxy groups and clusters 
that also have positive temperature gradients.
Radiating Bondi flows can also  
accommodate larger accretion rates ${\dot M}$ 
due to gas flowing in from an extended cluster environment.

To be useful in cosmological evolutionary computations 
and in interpreting accretion rates from X-ray observations, 
it is useful to keep our solutions 
as simple and general as possible. 
Consequently, our radiating 
steady solutions do not include spatially distributed mass
of stars or dark matter 
(e.g. Yahil \& Ostriker 1973; Cowie \& Binney 1977;
Fabian \& Nulsen 1977; Mathews \& Bregman 1978;
Quataert \& Narayan 2000;
Guo, Oh, \& Ruszkowski 2008), feedback energy, 
or rotation
(Park 2009; Brighenti et al. 2009; Narayan \& Fabian 2011).

When radiative losses are included, an initially hot gas
can sometimes cool to zero temperature before it reaches
the concentrated mass at the center,
particularly if the central mass is sufficiently small
or if there is no central gravitating mass at all.
This latter type of flow
toward a cooling site in the absence of gravity
experiences no tidal compression
so the gas temperature decreases continuously toward
the origin, becoming supersonic
before cooling to zero temperature.
Cooling site flows may describe the appearance and
evolution of cool, non-central low-entropy gas often observed
in galaxy clusters
far from the gravitating cluster core.

\section{Flow Equations}

Consider the equations for steady spherical Bondi accretion 
toward a point mass $M$ but also include 
optically thin radiative losses: 
\begin{equation}
{\dot M} = \rho u 4 \pi r^2
\end{equation}
\begin{equation}
\rho u {du \over dr} = - {dP \over dr} -  \rho { G M \over r^2}
\end{equation}
and
\begin{equation}
\rho u {d \varepsilon \over dr} =
{P \over \rho} u {d \rho \over dr}
- \left( \rho \over m_p \right)^2 \Lambda.
\end{equation}
The mass flow ${\dot M}$ is negative for inflowing gas, $u < 0$.
The specific thermal energy 
$\varepsilon = (P/\rho)/(\gamma - 1) = 3P/2\rho$ 
with $\gamma = 5/3$.
For simplicity we assume that the optically thin cooling
coefficient $\Lambda$ (erg cm$^3$ s$^{-1}$) is constant
throughout the flow.
While the temperature dependence of $\Lambda$ is well-known 
(e.g. Sutherland \& Dopita 1993), 
the quadratic density coefficient in the cooling term 
contributes much more to the non-linear cooling process 
than the comparatively slower variation of $\Lambda$ with 
temperature.
Moreover, a constant $\Lambda$ is easy to implement 
in checking gasdynamical hydrocodes.

The three flow equations can be reduced to two by eliminating
the density $\rho$ using equation (1):
\begin{equation}
{du \over dr} = {u \over r(3u^2 - 5 \theta)}
\left[ 10\theta -3 {GM \over r} + {2 \lambda \over u^2 r}
\right]
\end{equation}
\begin{equation}
{d\theta \over dr} = {\theta \over r(3u^2 - 5 \theta)}
\left[ -4u^2 + 2 {GM \over r}
+\left( 1 - {u^2 \over \theta}\right) {2 \lambda \over u^2 r}
\right]
\end{equation}
where 
\begin{equation}
\lambda = {{\dot M} \Lambda \over 4 \pi m_p^2}.
\end{equation}
and
\begin{equation}
P = \rho \theta ~~~{\rm where}~~~ \theta = { k T \over \mu m_p}
~~~{\rm so}~~~\varepsilon = {3\over 2} \theta.
\end{equation}

These equations become singular at the sonic radius 
$r = r_s$ where the flow speed equals the local 
adiabatic sound speed,  
\begin{equation}
u_s = c_s = (5 \theta_s/3)^{1/2}.
\end{equation}
The sonic radius is found by insisting that both square
brackets in the equations above vanish when
$u = (5 \theta/3)^{1/2}$, i.e.
\begin{equation}
r_s = {15 \theta_s GM - 6\lambda \over 50 \theta_s^2}.
\end{equation}
Use of L'H\^{o}spital's rule on the right side of either equation (4)
or (5) at the sonic point singularity gives the
following logarithmic slopes at the sonic radius 
(subscript $s$)
\begin{equation}
\left({d\log u \over d\log r}\right)_s =
{1\over2}[- b \pm (b^2 - 4c)^{1/2}]
\end{equation}
\begin{equation}
\left({d\log \theta \over d\log r}\right)_s =
-{2\over 3}\left[2 + {3 \lambda \over 5 \theta_s^2 r_s}
+ \left({d\log u \over d\log r}\right)_s \right]
\end{equation}
where 
\begin{equation}
b = 1 + {33 \over 100}{\lambda \over \theta_s^2 r_s}
~~~{\rm and}~~~
c = 1 + {39 \over 100}{\lambda \over \theta_s^2 r_s}
- {9 \over 40}{GM \over \theta_s r_s}.
\end{equation}

Beginning at the sonic point,
equations (4) and (5) can be integrated inward and outward
using the sonic slopes in equations (10) and (11). 
The sign option in equation (10) can be chosen for  
inflow or outflow. 
We restrict our discussion here to inflow solutions 
(minus sign).

The structure of solutions of equations (4) and (5) 
is more apparent when transformed to dimensionless variables 
\begin{equation}
\xi = {r \over r_s}~~~\eta = {u^2 \over {u_s}^2}~~~
\tau = {\theta \over \theta_s}
\end{equation}
in which the differential equations become 
\begin{equation}
{d\eta \over d\xi} = {2 \eta \over \xi(\eta - \tau)}
\left[ 2\tau - {30 \over A}{1 \over \xi} 
+ {12a \over A}{1 \over \eta \xi}\right]
\end{equation}
\begin{equation}
{d\tau \over d\xi} = {\tau \over \xi(\eta - \tau)}
\left[-{4\over 3}\eta 
+ {20\over A}{1 \over \xi} 
+ \left(1 - {5 \over 3}{\eta \over \tau}\right)
{12a \over A}{1 \over \eta\xi}\right]
\end{equation}
where $A = (15-6a)$ and
\begin{equation}
a = {1 \over 4\pi m_p^2} 
{\Lambda {\dot M} \over G M}{5 \over 3 c_s^2}.
\end{equation}
It is remarkable that 
the morphology of all flow solutions  
$\eta(\xi)$ and $\tau(\xi)$ depends on only one 
dimensionless parameter
\begin{equation}
a = 0.1008 
{({\dot M}/M_{\odot}{\rm~yr}^{-1}) 
(\Lambda/10^{-23}{\rm erg~cm}^{3}{\rm s}^{-1})
\over
(T_s/10^7{\rm K}) (M / 10^9 M_{\odot})}
\end{equation}
where $\mu = 0.61$. 
Since ${\dot M}<0$, $a$ must also be negative or zero and we 
consider values in the range 
$-\infty \le a \le 0$.

When radiative cooling is small, 
$\Lambda \rightarrow 0$, both $a$ and $\lambda$ also become 
small and equations (4) and (5) describe 
standard adiabatic Bondi flow.
When $\gamma = 5/3$, which we consider here,  
the traditional non-radiating Bondi accretion solution 
is unique, entirely subsonic and singular since 
the sonic point lies exactly at the origin, $r_s = 0$. 
Since our dimensionless variable $\xi = r/r_s$ 
is inappropriate when $r_s = 0$, we do not discuss 
the transition to standard adiabatic flow in detail here. 
When the central mass is small, 
$a \rightarrow -\infty$, solutions approach 
mass-free flow toward a cooling site with $M = 0$. 
For sufficiently large $|a|$ 
inflowing gas can cool completely ($\theta \rightarrow 0$)
before reaching the origin.

\section{Radiating Bondi Flows}

The top and central panels of Figure 1 
show eight superposed solutions 
of equations (14) and (15), $\eta(\xi)$ and $\tau(\xi)$, 
each corresponding to a different value of the parameter 
$|a| = 10^{-6}$, $10^{-2}$, $0.1$, 1, 1.5, 3, 10 and 100. 
The dotted extensions toward the left for each of these solutions 
describe supersonic flow within the sonic 
radius $\xi < 1$.
The dimensionless temperature $\tau = \theta/\theta_s$, 
shown in the central panel of Figure 1, 
has the largest variation with $|a|$.
For $|a| \ge 3$  
all plotted flows cool monotonically to zero temperature at 
a finite radius $\xi_{cool}$ not far within the sonic radius. 
At zero temperature the cooled gas would go into 
free fall toward the central mass, 
but this is not explicitly calculated. 

The parameter $a$ is related to ratios between the flow, 
free fall and cooling times
\begin{displaymath}
{t_{ff} \over t_{flow}} = {\pi \over 4}\alpha^{1/2}~~~
{t_{flow} \over t_{cool}} = {4 \over 3}{|a| \over \alpha}~~~
{t_{ff} \over t_{fcool}} = {\pi \over 3} {|a| \over \alpha^{1/2}}
\end{displaymath}
where $\alpha = 1 + (2/5)|a|$ and the times
\begin{displaymath}
t_{flow} = {r_s \over u_s}~~~
t_{ff} = {\pi \over 2}{{r_s}^{3/2} \over (2GM)^{1/2}}~~~
t_{cool} = {\varepsilon_s {m_p}^2 \over \Lambda \rho_s}
\end{displaymath}
are evaluated at the sonic radius $r_s$.

The dimensionless velocity variable 
$\eta(\xi) = u^2/u_s^2$ 
decreases monotonically with radius for all radiatively cooling 
solutions. 
For $|a| \le 10^{-6}$ 
the top panel in Figure 1 shows that the velocity variable  
$\eta(\xi)$ is nearly a single power law 
but becomes a double power law 
for $|a| \gta 0.1$, broken at the sonic radius $\xi = 1$.

Of particular interest are radiating flows  
with $|a| \gta 3$ in which the gas temperature 
drops to zero before it reaches the flow center.
These solutions are expected when the central 
mass $M$ (and/or the sonic point temperature $T_s$) 
becomes sufficiently small or when 
${\dot M}$ is unusually large.
When $|a| \ge 1$, 
the overlapping dotted lines for $\eta(\xi)$ in Figure 1 
obscure the different leftward terminations 
of $\eta(\xi)$ which do not extend 
within the radius $\xi_{cool}$ where $\tau \rightarrow 0$ 
in the bottom panel, 
with $\xi_{cool}$ increasing with $|a|$.

A temperature minimum at $\xi = \xi_{min}$ 
characteristic of radiating Bondi flows 
having smaller $|a|$ appears in Figure 1 
for the first time at $\log\xi_{min} \approx -1$
in the supersonic flow for $|a| = 1.5$. 
As $|a|$ decreases further, the location $\xi_{min}$ of the  
minimum in $\tau(\xi)$ moves progressively toward larger 
radii: $\log\xi_{min} \approx -0.1$, 0.6, and 1.4 for 
$|a| = 1$, 0.1 and 0.01 respectively.
For $|a| \le 10^{-6}$ the temperature minimum 
has moved to $\log\xi > 2$ and no longer appears in Figure 1.
For these very small values of $|a|$ 
the profiles for both $\eta(\xi)$ and $\tau(\xi)$ 
converge to power laws in $\xi \lta 100$ that 
appear as single straight lines in the region 
plotted in Figure 1.

Using equation (1), the corresponding
dimensionless gas density variable
\begin{equation}
\delta = {\rho \over \rho_s} = {1 \over \eta^{1/2}\xi^2}
\end{equation}
is normalized with the gas density at the
sonic radius $\rho_s$.
The bottom panel in Figure 1 shows that 
$\delta(\xi)$ decreases monotonically with radius 
for all solutions. 

The flow profiles in Figure 1 for velocity, temperature 
and density in dimensionless variables have morphologies 
that depend only on the single dimensionless
parameter $|a|$.
However, when these results are converted into physical units, 
(two of) the fundamental dimensional flow parameters in 
equation (16) -- 
$M$, ${\dot M}$, $T_s$ and $a$ -- 
become individually important,
particularly for the normalizing factors $r_s$ and $\rho_s$. 
The sonic radius 
\begin{equation}
r_s = {{(15 - 6a)}\over50}{GM \over \theta_s}
\end{equation}
depends on $GM/\theta_s$ as well as $a$ 
and the density at the sonic radius 
\begin{equation}
\rho_s = -{{\dot M} \over 4\pi}
{1 \over {r_s}^2}
\left({3 \over 5\theta_s} \right)^{1/2} 
\end{equation}
has its own unique dependence on flow parameters.

All flows with $|a| \lta 3$ have a central thermal peak. 
It is of interest to consider the total 
energy emitted in the central thermal peak $\xi < \xi_{min}$
\begin{equation}
L = \int (\rho/m_p)^2 \Lambda 4 \pi r^2 dr 
= L_{c} \int_0^{\xi_{min}} \delta^2 \xi^2 d\xi
\end{equation}
where
\begin{equation}
L_{c} = \left({3\over 5}\right)^{1/2}{1 \over 4 \pi {m_p}^2}
{50 \over (15-6a)}{{\dot M}^2 \Lambda \theta_s^{1/2}\over GM}.
\end{equation}
The asymptotic behavior of flow variables in the thermal peak 
as $\xi \rightarrow 0$ is: 
$\eta \propto \xi^{-1}$, $\delta \propto \xi^{-3/2}$ 
and $\tau \propto \xi^{-1}$. 
Consequently, the integral for the total peak luminosity,   
$\propto \int d\xi/\xi$, diverges logarithmically at the origin.
(If $\Lambda$ had a bremsstrahlung temperature 
dependence $\Lambda \propto T^{1/2}$, the divergence 
would be even stronger.)
Evidently this divergence is related to the 
concept of a point mass 
which can increase the internal energy and density 
of radially inflowing gas without limit.
Obviously rotation, magnetic or relativistic effects must 
eventually alter the nature of 
idealized radiating Bondi flows within some radius, 
removing the apparent divergence of the luminosity $L$ 
from the thermal peak.

In radiating Bondi 
flows having central thermal peaks the temperature 
in the peak 
increases as $\tau \propto \xi^{-1}$ but  
the dimensionless entropy factor, 
$\sigma = \tau/{\delta}^{2/3} \propto \xi^0$, 
is constant and the radiative cooling time is 
longer than the flow time.
Gas flows supersonically to the center along an adiabat.
But this adiabat differs from that of 
classical adiabatic Bondi accretion 
for $\gamma = 5/3$ in 
which gas flows subsonically toward the central point mass, 
becoming sonic as $r \rightarrow r_s = 0$.
In our radiating flows the rapidly converging and accelerating 
central flow becomes effectively adiabatic  
for all values of $a$ because of the short flow time, 
not because it stops radiating. 
Moreover, 
this central pseudo-adiabatic flow in $r < r_s$ 
is entirely supersonic, unlike the subsonic central flow 
in classical Bondi flow. 

The adiabatic character of asymptotic flow in 
central thermal peaks 
can be understood by comparing the two terms on the right 
in equation (3) that describe the importance of 
compression heating and radiation losses respectively. 
The ratio of thermal energy change due to 
radiation and compression is therefore
\begin{displaymath}
\sim {\Lambda {\dot M}  \over 4 \pi m_p^2}  
\cdot {1  \over \theta u^2 r} 
= {3  \over5 } 
\cdot {\Lambda {\dot M}  \over 4 \pi m_p^2 } 
\cdot {1  \over \theta_s^2 r_s } 
\cdot {1  \over \tau\eta\xi } 
\propto \xi.
\end{displaymath}
While the luminosity due to 
cooling by radiative losses diverges as $r \rightarrow 0$ 
during inflow in the thermal peak, 
the radiative term in equation (3) nevertheless becomes 
progressively smaller relative to the term for compressional heating 
which diverges even faster.
As $\xi \rightarrow 0$ 
the gas temperature is altered only by compression, 
explaining the adiabatic nature of the flow inside 
the central thermal peak.

\subsection{Cooling Site Flows}

Cooling site solutions in which 
$M \rightarrow 0$ are a well defined limit
of radiating Bondi flows. 
In this limit $a \rightarrow -\infty$ while 
${\dot M}$ and $\Lambda$ remain finite.
Flows for large $|a|$ in Figure 1
remain essentially unchanged
for $|a| \ge 100$, 
so the $|a| = 100$ profiles can be regarded  
as essentially equivalent to the 
(non-gravitating) cooling site flow. 
In cooling site flow the gas cools completely 
at some finite radius $\xi_{cool} < 1$ where all its entropy 
$\sigma = \tau/{\delta}^{2/3}$ 
is radiated away. 
In these flows the total radiated emission 
within the sonic radius 
\begin{displaymath}
L \approx L_{c} \log(1/\xi_{cool})
\end{displaymath}
is finite.

\section{Confronting Observations}

Of interest are potential sources of error when accretion 
rates are estimated by calibration with radiating and non-radiating 
Bondi flows. 
For this purpose it is necessary to convert 
from dimensionless to physical variables.

\subsection{Classical Bondi Flows}

In traditional Bondi accretion with $\gamma = 5/3$ 
gas flows with constant entropy toward a mass point 
at the origin. 
Very far from the mass point the gas is at rest with uniform 
density $\rho_{\infty}$ and temperature or sound speed $c_{\infty}$, 
that define the adiabat for the entire flow.
However, in Bondi solutions for a typical monatomic gas 
with $\gamma = 5/3$ the only flow from large radii 
to the origin must be fully subsonic, reaching the 
sound speed just at the origin, $r_s = 0$.
When $\gamma = 5/3$ the energy integral for 
the steady flow is 
\begin{equation}
{1 \over 2}u^2 + {3 \over 2}c_s^2 - {GM \over r}
= {1 \over 2}{u_{\infty}}^2 + {3 \over 2}c_{\infty}^2.
\end{equation}
If the gas velocity $u_{\infty}$ vanishes as $r \rightarrow \infty$, 
the Bondi accretion rate is
\begin{equation}
{\dot M} = \pi (GM)^2 {\rho_{\infty} \over c_{\infty}^3} 
\end{equation}
where the coefficient $\rho_{\infty}/c_{\infty}^3$ 
is related to the inverse of the entropy factor, 
$c_{\infty}^3/\rho_{\infty} \propto (T/\rho^{2/3})^{3/2}$. 
For a fixed central mass the accretion rate decreases with 
increasing gas entropy.
When estimating the classical quadratic Bondi dependence,
${\dot M}\propto M^2$, 
the entropy factor $c_{\infty}^3/\rho_{\infty}$
in equation (24) is assumed to be independent of $M$.
But it is easy to imagine  astronomical
correlations between entropy and $M$ that are unrelated to
the hydrodynamics: larger black hole masses 
are often found at the centers of more massive dark halos 
containing gas with higher entropy.

The Bondi accretion rate ${\dot M}$ onto galaxy or cluster-centered
black holes can be estimated 
from X-ray observations of gas temperature 
and density profiles. 
Gas observed at some radius just larger than 
$\sim GM/{c_s}^2$, where the black hole potential begins 
to dominate, is assumed to be essentially at rest  
For example, Allen et al. (2006) estimated  
$\rho_A$ and $T_A$ from (inward extrapolated) observations 
at radius $r_A = GM/2{c_A}^2$ where 
${c_A} = (\gamma kT_A/\mu m_p)^{1/2}$ is the local sound speed 
at $r_A$ and $M$ is known from observations of the 
central galaxy. 
Since the entropy factor ${c_s}^3/\rho$ is constant 
in Bondi flow, it can be observed at any radius.
However, the Bondi flow morphology and equation (24) 
require $u_{\infty} = 0$.


\subsection{Radiating Bondi Flows}

In radiating Bondi flows radiation losses 
occur at all radii so the flow at large distances 
from the gravitating mass point 
necessarily has gradients unlike the traditional Bondi flow. 
For $\xi \gta 100$,
$d\log \eta/d\log \xi = -2$ and $d\log \tau/d\log \xi = 1$
are excellent approximations to radiating solutions 
in Figure 1 when $\xi \gta \xi_{min}$. 
Consequently, 
the sound speed and gas density at infinity are not defined 
nor is this limit relevant in comparing with observation.

Radiating accretion flows are inherently non-adiabatic.
However, for $|a| \lta 1.5$ radiating supersonic Bondi flows 
ultimately become adiabatic as $\xi \rightarrow 0$ 
in the central thermal peak, having 
passed through a temperature minimum and a sonic point.
The sonic radius,
\begin{equation}
r_s = 13 {(15-6a) \over 21}
{(M / 10^9 M_{\odot}) \over (T_s/10^7{\rm K})}~~{\rm pc}
\end{equation}
is likely to be small, not directly observable.
Remarkably, the morphology of all radiating Bondi flows 
is determined by a 
single dimensionless parameter $a$. 
The accretion rate is 
\begin{equation}
{\dot M} = a \cdot 4\pi m_p^2 {3 c_s^2 \over 5} 
{G M \over \Lambda}
\end{equation}
where $c_s = u_s$ is the sound speed at the sonic radius.

Alternatively, 
by using equation (19) the radiating accretion rate
can be written in terms of the sonic radius
\begin{equation}
{\dot M} = a \cdot 4\pi m_p^2
{G M \over \Lambda}
\left( 15 - 6a \over 50 \right) {GM \over r_s}.
\end{equation}
While the quadratic mass dependence
in this less useful representation ${\dot M} \propto M^2$
resembles the classical Bondi relation,
its coefficient $a(15-6a)/r_s$ cannot be directly observed and
has a complex dependence on the flow variables.

Flows having the same ${\dot M}/M$ and $T_s$ (and therefore
$a$) have identical temperature and density profiles
$T(r)$ and $\rho(r)$ and for these 
flows the linear relation ${\dot M} \propto M$ holds exactly.
However, the relation ${\dot M}\propto aM{c_s}^2$ 
from equation (26) 
does not in general imply that ${\dot M}\propto M^p$ 
with $p = 1$, nor does it imply that ${\dot M}$ and $M$ are 
necessarily related with a simple power law. 
While accretion flows expressed in dimensionless variables, 
as in Figure 1, depend on the single parameter $a$,
the same flows translated into dimensional variables 
$u(r)$, $T(r)$ and $\rho(r)$ depend on any two of the
parameters $a$, ${\dot M}$ and $T_s$, assuming that
$M$ is independently known from observations of the central galaxy.

A more general relationship between ${\dot M}$ and $M$ 
can be established by considering intersecting flows.
In particular, by varying ${\dot M}$ and $T_s$
it is possible to construct radiating flows having different
central masses $M$ that share the same temperature $T$ and
density $n_e = \rho/(1.17m_p)$ at the same radius $r$.
Such intersecting flows have different $a$ and different
morphologies.
For example, Figure 2 shows three radiating 
Bondi flows that intersect with identical gas 
properties. 
Only the subsonic regions of these flows are plotted, 
extending from the sonic radius $r_s$ at the left to
1 kpc at the right. 
The solid line profile is a reference solution with parameters
$M = 3\times 10^9$ $M_{\odot}$,
${\dot M} = 0.3$ $M_{\odot}$ yr$^{-1}$, and
$T_s = 3\times 10^6$. K.
The dashed lines are two radiating Bondi flows
with lower central mass $M/3$ that 
intersect the reference flow at points 1 and 2 
where they have the same $T$ and $n_e$ at $r_1$ and $r_2$. 
The upper dashed flow profiles $(M/3)_1$ 
in both panels of Figure 2 intersect at 
$r_1 = 152.2$ pc (point 1) 
where $T = 2.21\times 10^6$ K and $n_e = 0.182$ cm$^{-3}$.
The parameters of this flow are
$M = 10^9$ $M_{\odot}$,
${\dot M} = 0.07917$ $M_{\odot}$ yr$^{-1}$, and
$T_s = 1.76\times 10^7$ K.
The lower dashed flow profiles $(M/3)_2$ 
in Figure 2, based on parameters
$M = 10^9$ $M_{\odot}$,
${\dot M} = 0.17929$ $M_{\odot}$ yr$^{-1}$, and
$T_s = 9.00\times 10^5$ K, 
intersect at $r = 624.3$ pc (point 2) where
$T = 1.432\times 10^6$ K and $n_e = 0.03568$ cm$^{-3}$.

Of interest are the differing values of ${\dot M}$ and $M$ 
for the two flows involved at each of these intersections 
and in particular the mean power law variation
${\dot M} \propto M^p$ that approximately 
describes how the accretion rate varies with central 
mass for flows with central masses in the range 
$M/3 - M$, all of which intersect with the 
reference solution at the same point. 
From the two very specific intersections shown in Figure 2  
at points 1 and 2 
we find that ${\dot M} \propto M^p$ with 
$p = 1.21$ and $0.47$ respectively.
The exponent $p$ differs from unity 
because the three intersecting flows 
in Figure 2 have different overall
morphologies and values of $a$.
The three flows in the upper 
panel of Figure 2 can be ordered -- 
$(M/3)_1$, $M$, $(M/3)_2$ -- with decreasing
$r_{min}/r_s$ where $r_{min}$ is the radius 
of minimum gas temperature and the sonic 
radius $r_s$ is defined by the leftmost extent 
of each flow.
A qualitatively similar morphological sequence 
for decreasing $r_{min}/r_s = \xi_{min}$ is 
apparent in the central panel of Figure 1 for 
flows with $|a| = 0.01$, 0.1 and 1.
This is consistent with similarly increasing values of 
$|a| = 0.0137$, 0.1008 and 0.6027 
respectively that characterize the intersecting 
flows $(M/3)_1$, $M$, and $(M/3)_2$ in Figure 2.

The dotted lines $(2M/3)_1$ in Figure 2 show a third root 
through point 1 with parameters: $2M/3 = 2\times 10^9$ $M_{\odot}$,
${\dot M} = 0.1725$ $M_{\odot}$ yr$^{-1}$, 
$T_s = 9.3\times 10^6$ K and 
$|a| = 0.02806$.
The variation of $p$ between the three roots is 
not constant: $p = 1.12$ for $(M/3)_1 \rightarrow (2M/3)_1$ and 
$p = 1.36$ for $(2M/3)_1 \rightarrow M$.
Evidently, there is no general power law relation 
${\dot M} \propto M^p$ that applies to all intersecting 
flows of this type, although local values of 
$p$ are not far from unity.

Also clear from Figure 2 
is that the mass accretion rate ${\dot M}$ 
cannot be determined 
by observations of $T$ and $n_e$ at a single radius. 
The parameters of radiating flows, ${\dot M}$ and $T_s$,  
must be found with fits to a spatially extended 
set of observations 
that extend close to or within the radius of 
minimum temperature.
The best-fitting flow parameters ${\dot M}$ and $T_s$ 
can be found with iterative integrations 
out from the sonic radius as we have done here.

Three additional points:
(1) Although we consider intersecting accretion flows 
having different $M$ 
to explore the ${\dot M} \propto M^p$ relation, it must 
be stressed that not all flow solutions intersect 
nor do they intersect as shown in Figure 2. 
For example, subsonic entropy factor profiles 
$S(r) = T(r)/{n_e}^{2/3}(r)$
do not intersect when only the sonic temperature $T_s$ 
is varied.
Subsonic entropy profiles in which only ${\dot M}$ 
varies typically intersect but, 
unlike Figure 2, the values of $T$ and $n_e$ are 
not identical at intersections having the same entropy 
$T/{n_e}^{2/3}$.
The unusual flow intersections in Figure 2 
with identical $T$ and $n_e$ 
but varying mass $M$, $T_s$ and ${\dot M}$ 
reveal a relationship between $a = a(T_s,{\dot M}/M)$ 
and the exponent $p$.
Unlike traditional Bondi flows, 
the ${\dot M}(M)$ relation for radiating Bondi flows is not 
defined by the entropy $T/{n_e}^{2/3}$ at a single 
radius of observation.
(2) When $M$ is known in advance, 
the morphology parameter depends on the sonic temperature 
and accretion rate, $a = a(T_s,{\dot M})$.
As $T_s$ and ${\dot M}$ are varied to match observations 
of a single flow or when a group of similar flows are 
compared, the range of $T_s$ and ${\dot M}$ for successful 
fits is limited 
and correspond to a relatively small variation in $a$. 
When $M$ and ${\dot M}$ are compared among these flows, 
we expect $p \approx 1$.
(3) We draw attention to the relative insensitivity 
of the index $p$ to the morphology parameter $a$ 
in the intersecting examples illustrated in Figure 2. 
As  $|a|$ increases from 0.0137 to 0.6027, a factor of 44, 
$p$ changes only from 1.21 to 0.47.
In addition, for massive galaxy-centered black holes 
$a$ is constrained by observations:
$|a| \lta 3$ is required to avoid total off-center cooling 
(eqn 28)  
which is not generally observed and $|a|$ cannot be 
very much less than 0.01 
since temperature minima are rarely observed 
at large radii.
This also suggests that $p$ lies in a restricted region 
near unity. 

In Figure 3 we compare the accretion rate for the 
reference solution ${\dot M} = 0.3$ $M_{\odot}$ yr$^{-1}$
with estimates of ${\dot M}$ using the classical Bondi 
procedure (eqn 24). 
The Bondi mass flow rates ${\dot M}$ are determined 
from flow local gas temperature and density 
of the reference solution.
It is seen that the classical Bondi solution 
considerably underestimates ${\dot M}$ in the 
region plotted. 
The decreasing classical Bondi ${\dot M}$ with radius 
mirrors the monotonically 
increasing entropy in the radiating solution.

Equation (24) for Bondi flow 
is used extensively in computations of cosmological 
black hole accretion 
(e.g. Springel et al. 2005; 
Johansson et al. 2009;
Dubois et al. 2010; 
Kim et al. 2011).
But successful black hole growth is only possible 
if the Bondi accretion rate is enhanced by a large 
{\it ad hoc} dimensionless coefficient, 
while such a factor $a$ appears naturally in radiating 
Bondi flows (eqn 26).
Bondi accretion is also invoked to estimate accretion 
rates ${\dot M}$ from X-ray observations 
of hot gas surrounding galaxy-centered and cluster-centered 
black holes 
(e.g. Allen et al. 2006;
Balmaverde et al. 2008;
Vattakunnel et al. 2010).
However, as with the cosmological computations, 
X-ray observations of sources having 
greater feedback power require mass accretion rates considerably 
in excess of the Bondi rate 
(Rafferty et al. 2006;
Hardcastle et al. 2007;
McNamara et al. 2011). 
In addition to the large mass accretion rates implied by 
radiative emission from luminous 
black holes in AGNs and quasars, 
a comparable accretion rate is required to 
provide the huge mechanical power 
expended by cluster-centered black holes 
in expanding the cluster gas to match observed 
gas mass fraction profiles 
(Mathews \& Guo 2011).


In radiating Bondi flows 
the temperature dependence in this outer region 
$\xi > \xi_{min}$, $\tau \propto \xi$, 
is a reasonable match to gas temperature profiles in the 
central region of most galaxy groups and clusters.
Moreover, the density in $\xi > \xi_{min}$ varies 
asymptotically as $\delta \propto \xi^{-1/2}$ 
so the entropy is not constant 
but varies rather significantly with 
radius, $\sigma = \tau/\delta^{2/3} \propto \xi^{4/3}$,  
at large $\xi$ in radiating Bondi flows.
However, hot gas observed in more isolated elliptical galaxies 
can have $dT/dr < 0$.  
This can be understood if the gas is confined by a halo of lower 
mass (Humphrey et al. 2006) or, at smaller radii, 
by compressional heating near the central black hole 
(Humphrey et al. 2008).

Positive temperature gradients typically observed 
on kpc scales in galaxy groups and clusters are 
evidently imposed by entropy-losing 
cooling flows in the gravitational  
potential of the stars and the group/cluster dark matter 
on larger multi-kpc scales.  
These flows on larger scales have local enclosed masses 
greatly exceeding that of the central black hole 
and contain gas with a larger virial temperature.
Consequently, the gas flow rate ${\dot M}$ 
toward cluster-centered black holes 
depends not just on the black hole mass 
but on the huge mass of baryonic gas bound to the dark halo 
of the surrounding galaxy group or cluster.
These large scale temperature and density gradients 
further confound estimates of the black hole 
accretion rate using the traditional Bondi recipe (eqn 24).
Once the innermost resolvable (or extrapolated)
gas density and temperature are 
interpreted as $\rho_{\infty}$ and $T_{\infty}$, 
the Bondi formula provides a unique local value of 
${\dot M} = \pi (G M)^2 \rho_{\infty}/c_{\infty}^3$ for each $M$. 
By contrast, the corresponding relation 
for radiating Bondi flows, ${\dot M} \propto a M T_s$, 
contains the additional 
dimensionless parameter $a$ that can 
be adjusted to fit much larger mass flows ${\dot M}$ 
that approach the central black hole 
from radiating gas bound to the 
surrounding group or cluster halos. 

Large values of the parameter 
$|a|$ can be regarded as a measure 
of large accretion rates that may cool significantly 
before reaching the central black hole. 
According to Figure 1, 
off-center cooling requires $|a| \gta 3$ 
which, with equation (26), corresponds to 
larger accretion rates  
\begin{equation}
|{\dot M}| \gta 30 
\left({\Lambda \over 10^{-23}{\rm erg~cm}^{3}{\rm s}^{-1}}\right)^{-1}
{T_s \over 10^7{\rm K}} 
{M \over 10^9 M_{\odot}}~{M_{\odot} \over {\rm~yr}}.
\end{equation}
This is roughly consistent with observations
of significant central star-formation in many 
X-ray luminous cool-core clusters 
(e.g. Donahue et al. 2010), 
although this occurs on scales larger than the sonic radius 
for the black hole.
This inequality also expresses a biomdality 
between hot 
($|a| \lta 3$; $|{\dot M}| \lta 30$ $M_{\odot}$ yr$^{-1}$) 
and cold ($|a| \gta 3$; $|{\dot M}| \gta 30$) 
modes of accretion by central black holes.
Radiating accretion flows that cool before reaching 
the central mass are those with small central mass $M$, 
lower temperature, or, as in the example above, 
large accretion rates ${\dot M}$
(Sarazin \& White 1987; 
Quataert \& Narayan 2000).

\section{Conclusions}

Gas flowing near accreting black holes can only be observed 
with radiation that necessarily carries entropy away from the flow, 
in violation of the adiabatic assumption inherent 
in traditional Bondi flows. 
However, when radiative cooling is included, 
the accretion flow solutions are qualitatively different, 
not merely perturbations or limiting cases of the 
original Bondi flow. 
Nevertheless, radiating solutions have 
widely varying morphologies that depend on a single 
dimensionless parameter and they all must pass through 
a sonic point.
Although the gas density (entropy) decreases (increases) 
monotonically with radius, 
many solutions exhibit a broad temperature minimum. 
Far from the mass point the temperature and entropy 
increase, allowing fits to be made to 
gas flowing subsonically in from a deeper potential well 
having a larger virial temperature.
Some radiating accretion flows do not pass through 
a temperature minimum, but instead cool completely before  
reaching the central mass. 
In ``cooling site'' flows, in which there is 
no central mass, total cooling occurs at a non-zero radius 
within the sonic point.

Of particular interest is the failure of traditional 
Bondi flows to accurately relate the mass accretion 
rate to the central mass.
The traditional Bondi result ${\dot M} \propto M^p$
with $p = 2$
is certainly incorrect for radiating flows. 
The examples we consider
indicate much smaller values $p \sim 1$ in which 
the precise value of $p$ is somewhat sensitive to the flow 
morphology.

The idealized radiating Bondi flow we describe here, while
certainly an 
improvement over the classic adiabatic Bondi flow, 
cannot be valid within some radius.
Any small rotation in the initial gas must ultimately 
flow into a disk. 
Nevertheless, these idealized spherical radiating 
solutions may provide a useful 
framework for understanding more realistic flows 
that include rotation, magnetic fields, viscosity, 
feedback, etc.
They may also be useful in evaluating 
the performance 
of central or non-central cooling in numerical computations. 

\acknowledgements
Our colleague Fabrizio Brighenti 
provided helpful advice.
We also acknowledge with thanks insights provided by 
the referee who challenged us to provide explicit 
examples of non-quadratic relations 
between ${\dot M}$ and $M$.
Studies of feedback gasdynamics in hot intracluster gas 
at UC Santa Cruz are supported by NSF and NASA grants
for which we are very grateful.


\clearpage

\begin{figure}
\vskip1.in
\centering
\includegraphics[width=4.in,scale=0.8,angle=0]{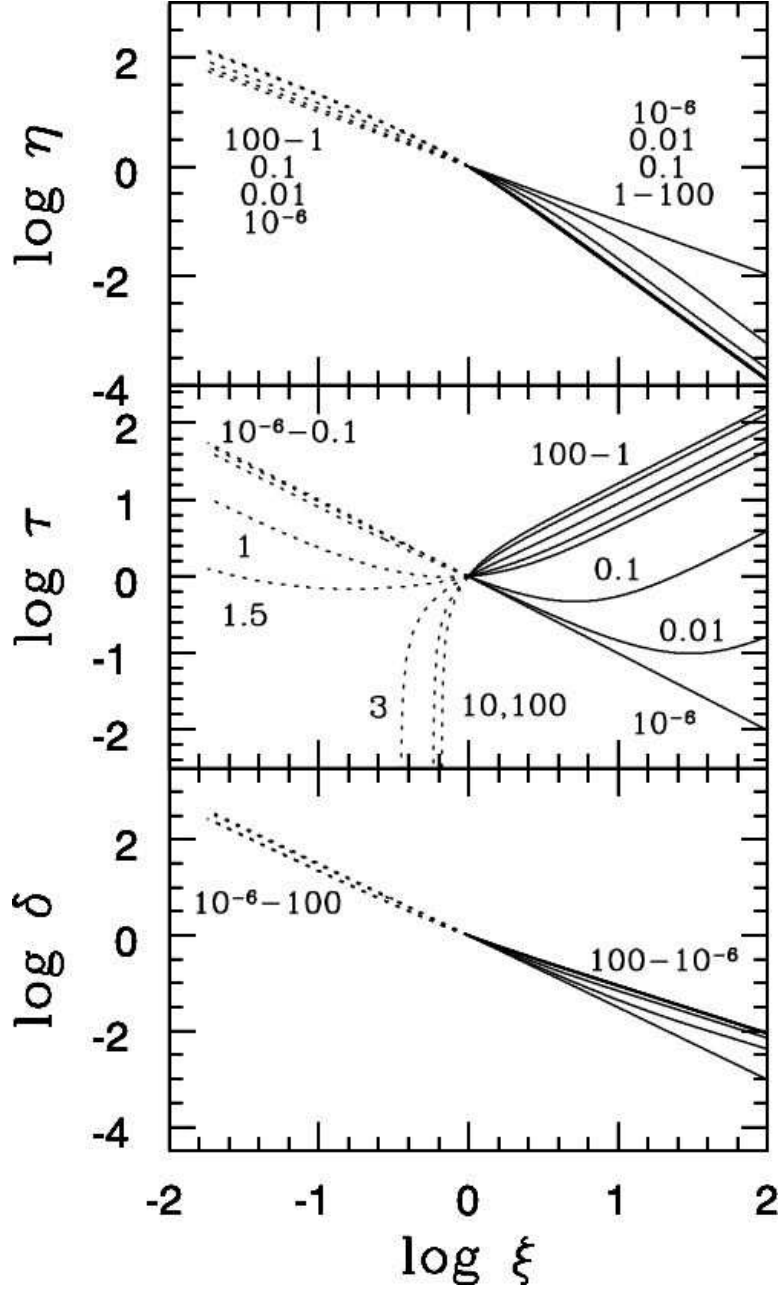}
\caption{
Plot of radiating Bondi flows in dimensionless variables.
{\it Top:} Dimensionless (square of the) velocity $\eta(\xi)$. 
{\it Center:} Dimensionless temperature $\tau({\xi})$.
{\it Bottom:} Dimensionless density $\delta({\xi})$.
Where possible, 
each curve is labeled with its dimensionless parameter $|a|$. 
Profiles that are too crowded to label individually 
are identified with the range in $|a|$ 
in order of decreasing variable $\eta$, $\tau$ 
or $\delta$.
}
\label{f1}
\end{figure}

\clearpage

\begin{figure}
\vskip1.in
\centering
\includegraphics[width=4.in,scale=0.8,angle=0]{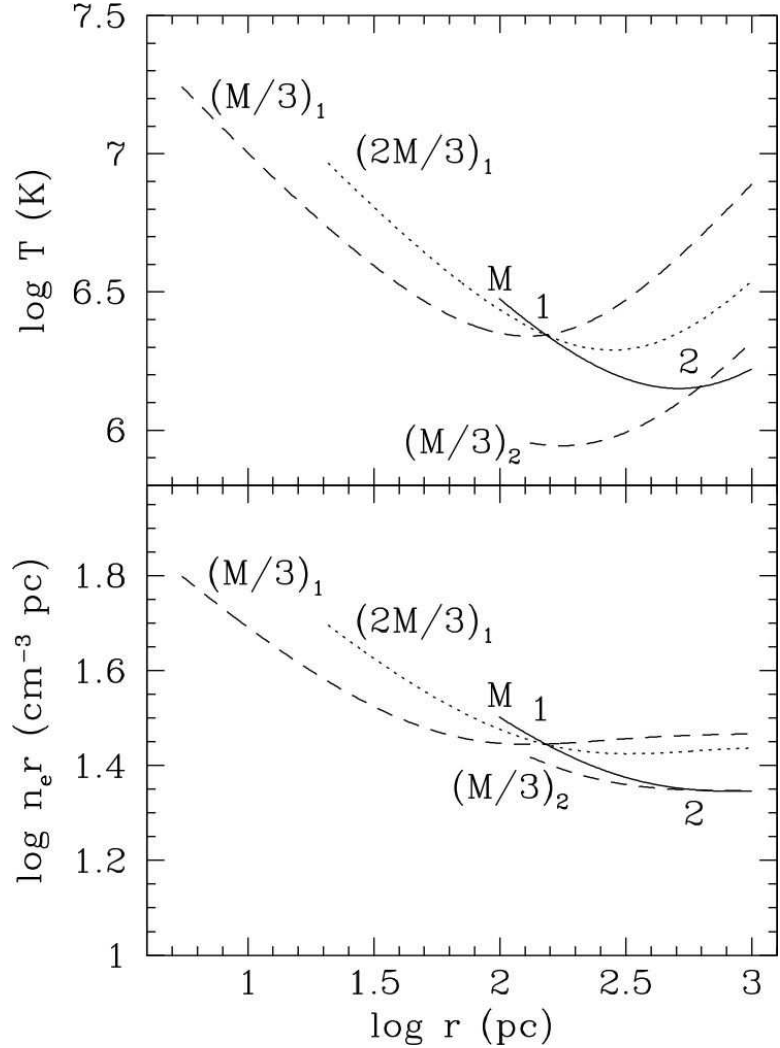}
\caption{
{\it Upper} and {\it Lower} panels show respectively 
gas temperature and density profiles of the subsonic 
regions of three radiating accretion flows, 
each extending from 1 kpc at the right 
to the sonic radius $r_s$ at the left.
For better visibility the gas density is multiplied 
by the radius in the lower panel.
Solid lines show a reference solution for which 
$M = 3\times 10^9$ $M_{\odot}$, 
${\dot M} = 0.3$ $M_{\odot}$ yr$^{-1}$, and
$T_s = 3\times 10^6$. K.
The dashed lines show two additional flows 
$(M/3)_1$ and $(M/3)_2$ for which 
the black hole mass $M = 10^9$ $M_{\odot}$ is three times 
smaller.
Parameters ${\dot M}$ and $T_s$ for the two dashed profiles 
have been chosen so that 
the intersections with the reference flow at points 1 and 2
at radius $r$ have the same temperature and density 
as the reference solution.
The dotted lines $(2M/3)_1$ show a third root through point 1 
for an intermediate mass.
The objective is to explore how ${\dot M}$ varies 
with $M$ for flows observed having identical 
$T$ and $n_e$ at some radius.
}
\label{f1}
\end{figure}

\clearpage

\begin{figure}
\vskip1.in
\centering
\includegraphics[width=4.in,scale=0.8,angle=0]{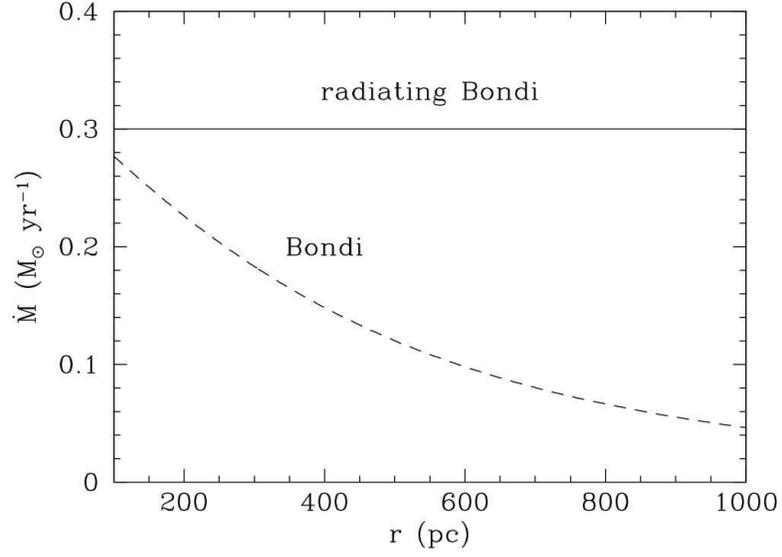}
\caption{
The solid line shows the constant mass accretion profile  
${\dot M} = 0.3$ for the reference radiating Bondi solution for 
$M = 3\times 10^9$ $M_{\odot}$,
${\dot M} = 0.3$ $M_{\odot}$, and
$T_s = 3\times 10^6$. K.
The dashed line shows the mass accretion rate 
for classic Bondi flows estimated with 
${\dot M} = \pi (GM)^2 {\rho_{\infty} / c_{\infty}^3}$ 
with the usual assumption 
$\rho_{\infty} = \rho(r)$ and $c_{\infty} = c_s(r)$.
}
\label{f1}
\end{figure}

\end{document}